\documentclass[nofootinbib,amsmath,prd,aps,tightenlines,preprint,preprintnumbers,letterpaper,12pt]{revtex4}

\usepackage{mymacros}
\usepackage{slashed}
\usepackage{verbatim}
\usepackage{tabularx}

\usepackage{color}
\definecolor{lcolor}{rgb}{0,0.0,0.0}

\usepackage[bookmarks=false,
	colorlinks=true,
	linkcolor = black,
	urlcolor = lcolor,
	citecolor = lcolor]{hyperref}            
\usepackage{graphicx}
\usepackage{epstopdf}

\usepackage{bm}
\usepackage{epsfig}
\usepackage{graphics}
\usepackage{xspace}
\usepackage{slashed}
\usepackage{feynmp}
\usepackage{ams fonts}
\usepackage{sidecap}
\usepackage{dcolumn} 
\usepackage{longtable} 
\usepackage{booktabs}
\usepackage{color}
\usepackage{float}

\definecolor{green}{rgb}{0,.5,0}

\hypersetup{pdfpagemode=FullScreen}

\begin{document}
\setlength\baselineskip{17pt}

\title{Perturbative Renormalization and Mixing of Quark and Glue Energy-Momentum Tensors on the Lattice}
\vspace{0.3 in}

  \author{Michael J. Glatzmaier\footnote{Electronic address: \href{mailto:michael.glatzmaier@gmail.com}{michael.glatzmaier@gmail.com}}}
  \author{Keh-Fei Liu\footnote{Electronic address: \href{mailto:liu@pa.uky.edu}{liu@pa.uky.edu}}}
  \author{Yi-Bo Yang\footnote{Electronic address: \href{mailto:ybyang@pa.uky.edu}{ybyang@pa.uky.edu}}}
  \affiliation{Department of Physics and Astronomy, \& Center for Computational Sciences, University of Kentucky, Lexington, KY 40506}
 
\date{\today\\ \vspace{1cm} }

\begin{abstract}
We report the renormalization and mixing constants to one-loop order for the quark and gluon energy-momentum (EM) tensor operators on the lattice.  A unique aspect of this mixing calculation is the definition of the glue EM tensor operator.  The glue operator is comprised of gauge-field tensors 
constructed from the overlap Dirac operator.  The resulting perturbative calculations are performed using methods similar to the Kawai approach using the Wilson fermion and gauge actions for all QCD vertices and the overlap Dirac operator to define the glue EM tensor.  Our results are used to connect the lattice QCD results of quark and glue momenta and angular momenta to the $\overline{\text{MS}}$ scheme at input scale $\mu$. 
\end{abstract}
\maketitle

\section{Introduction}
The nucleon spin problem is still an outstanding issue in QCD.  The problem originated from the European Muon Collaboration (EMC) experiment which indicated that the contribution of the quark spin to the proton spin was only 25\% of the theoretical prediction in the quark model.  To settle this issue, a more precise determination of both the quark and glue contributions to the nucleon spin are necessary.  But in addition to the increased experimental precision, it is a difficult issue to address theoretically as well.  In this regard, lattice determinations of the momentum and angular momentum are indispensable.  

Recent lattice calculations of the quark orbital angular momenta in the connected insertion have been carried out for the connected insertions~\cite{Mathur:1999uf,Hagler:2003jd,Bratt:2010jn,Brommel:2007sb,Syritsyn:2011vk,Sternbeck:2012rw,Alexandrou:2013joa,Alexandrou:2016mni}, and it was shown to be small in quenched calculations~\cite{Mathur:1999uf} and near zero in dynamical fermion calculations~\cite{Hagler:2003jd,Bratt:2010jn} due to the cancellation between the $u$ and $d$ quarks. The disconnected insertion contribution is also investigated on the lattice using dynamical fermions but the signal is noisy~\cite{Abdel-Rehim:2013wlz}.  The Gluon helicity distribution $\Delta G(x)/G(x)$ from COMPASS and STAR experiments was found to be close to zero~\cite{Stolarski:2010zz,Djawotho:2013pga} while the evidence of a non-zero $\Delta G(x)$ is confirmed recenetly~\cite{deFlorian:2014yva,Nocera:2014gqa}.  Additionally, it has been argued based on analysis of single-spin asymmetry in unpolarized lepton scattering from a transversely polarized nucleon that the glue orbital angular momentum vanishes~\cite{Brodsky:2006ha}, leaving us a in a `Dark Spin' scenario.

A full lattice calculation of the quark and glue momenta and angular momenta has just been completed with quenched Wilson fermion and gluon actions, where both the quark connected and disconnected insertions are included~\cite{Deka:2013zha}.  In combining with earlier work on the quark spin, a result for the quark orbital angular momentum was obtained.  It was found that the $u$ and $d$ quark orbital contributions indeed largely cancel in the connected insertion, as in the dynamical fermion calculation~\cite{Hagler:2003jd,Bratt:2010jn}, however their contributions in the disconnected insertion, including the strange quark, are on the order of 50\% of the total nucleon spin. Even though the glue momentum in the proton has been studied in serval recent works~\cite{Horsley:2012pz,Alexandrou:2013tfa}, the glue angular momentum was obtained for the first time with the gauge field strength tensor for the glue operators defined by the overlap Dirac operator. 

Our aim in this paper is to calculate the renormalization and mixing constants necessary to extract continuum physics from a lattice calculation of the quark and glue angular momentum operators.  These one-loop Z-factors calculated from lattice perturbation theory are a crucial ingredient in computing the matching conditions between lattice calculations, which are regulated with an explicit lattice spacing `$a$', and experimental results, which are quoted in the $\overline{\text{MS}}$ scheme.  As the one-loop perturbative calculations involving the overlap Dirac operator are lengthy, we have written several scripts in Mathematica and {\tt python} to carry out the calculation analytically as far as possible.  At the end of all manipulations, a final series of numerical integrations is necessary before quoting the renormalization constants.  The quark sector of this calculation follows closely the calculations in~\cite{Capitani:1994qn}, and so the finite pieces of these results have been relegated to the appendices of this work.  The glue sector however is new and the finite pieces of those diagrams involving the glue angular momentum operator $Z_{G\to Q}$ and $Z_{G\to G}$ have been listed in the conclusion.
  
We have organized this paper as follows, in section \ref{formalism}, we outline the general aspects of the mixing calculation and highlight terminology used for the remainder of the paper.  In section \ref{operators} we sketch the derivation of the Feynman rules used for the glue EM tensor operator defined from the overlap Dirac derivative and give the details in section~\ref{operators} as well as Appendix \ref{appendixA}.  In section~\ref{renormalization} we present the renormalization conditions used and in section \ref{methodology}, we detail our approach in extracting the finite contributions to the renormalization constants.  We present our results for each calculation in section~\ref{results}. We conclude and summarize our goals for future work in section \ref{conclusions}.   
   
 \section{Formalism}\label{formalism}
\noindent
The QCD angular momentum operators are defined according to the generators of the Lorentz transformation~\cite{Ji:1996ek}
\bea
 J^i &\equiv& \frac{1}{2}\epsilon^{ijk}\displaystyle\int d^3 x M^{0jk} (\vec{x}),
 \label{am_operator}
\eea
where \ $M^{0ij}$ is the angular momentum density,
\bea
  M^{\alpha\mu\nu} (x) 
  &=& T^{\alpha\nu} x^\mu - T^{\alpha\mu} x^\nu,
\eea
and here, $T^{\mu\nu}$ is the symmetric, gauge-invariant, QCD energy-momentum tensor.  

One can then decompose the energy momentum tensor into a gauge-invariant sum of its quark and glue contributions,
\bea
T^{\mu\nu} = T^{\mu\nu}_q + T^{\mu\nu}_g,
\label{energy_tensor}
\eea
where the subscripts, $q$ and $g$, stand for the quark and glue operators, respectively.  Explicitly, these operators are equivalent to the leading twist operators in unpolarized DIS in Euclidean space,
\bea
T^{\mu\nu}_q = \frac{1}{4}\mathcal{S}\>\sum_f\bar{\psi}_f\left( \gamma^{\mu}  \overrightarrow{D}^\nu - \gamma^\mu  \overleftarrow{D}^\nu\right )\psi_f,
\label{quark_piece}
\eea
where $\mathcal{S}$ denotes that $T^{\mu\nu}$ is symmetrized with respect to indices $\mu$ and $\nu$ and $f$ denotes quark flavor.  For the glue operator,
\bea
T^{\mu\nu}_g = \frac{1}{2}\mathcal{S}\>G^{\mu\alpha}G^{\nu}_\alpha,
\label{gluon_piece}
\eea
where a trace over color indices has been suppressed, and $G$ denotes the gauge field strength tensor.  These equations allow one to write $\vec{J}$ as a gauge invariant sum,
\bea
\vec{J}_{\text{QCD}} = \vec{J}_g + \vec{J}_q,
\label{total}
\eea
where, using Eq.(\ref{am_operator}), the $i^{\text{th}}$ component of $J$ is,
\bea\label{ji_gauge_invariant}
J^i_{q,g} = \frac{1}{2}\epsilon^{ijk}\int\>d^3x \left(T^{0k}_{q,g}x^j-T^{0j}_{q,g}x^k\right).
\eea
\noindent
One can also re-express $J_q$ and $J_g$ into a form more suitable for physical interpretation using the QCD equations of motion~\cite{Ji:1996ek,Hoodbhoy:1988am}, one arrives at the well known result,
\bea
  \vec{J}^q &=& 
  \int d^3x \,\frac{1}{2}\, 
  \bigg{[} \overline\psi\,\vec{\gamma}\,\gamma^5 \,\psi 
  \label{gi_quark_mom}
 + \psi^\dag \, \left(\vec{x} \times (i \vec{D})\right)  \,\psi \bigg{]}, \\
 \vec{J}^g &=& \int d^3x \,\bigg{[} \vec{x} \times ( \vec{E} \times \vec{B} )\bigg{]}.
\label{gi_gluon_mom}
\eea
where both the color and flavor indices are suppressed. The first term of Eq.~(\ref{gi_quark_mom}) is identified as the quark spin operator 
$\frac{1}{2}\>\vec\Sigma^q$ and the second term as the orbital angular momentum 
operator ($\vec{L}^q$).\ Thus, we write the total angular momentum for quarks,
\bea
\vec{J}^q &=&  \frac{1}{2} \vec\Sigma^q + \vec{L}^q.
\label{quark_ang_op_def_split_2}
\eea
%
%
Collecting the results found in Eqs.~(\ref{total}),\ (\ref{gi_quark_mom}) and 
(\ref{gi_gluon_mom}),\ the angular momentum operator in QCD can be 
expressed as a gauge-invariant sum~\cite{Ji:1996ek},
\bea
  \vec J_{\text{QCD}} = \vec J^q + \vec J^g
  = \frac{1}{2} \vec\Sigma^q + \vec{L}^q + \vec{J}^g.
\label{ang_op_def_split_2}
\eea
One must measure all the three quantities in Eq.~(\ref{ang_op_def_split_2}) on the lattice in order to address the `Dark Spin' scenario from first principles.  The first term appearing in Eq.(\ref{gi_quark_mom}) measures the quark spin contribution to the proton spin and several studies have already computed this operator on the lattice, the details can be found in~\cite{Dong:1995rx, Fukugita:1994fh, Gusken:1999xy} and the recent updates on the disconnected contributions can be found in Refs.~\cite{Babich:2010at,QCDSF:2011aa,Engelhardt:2012gd,Abdel-Rehim:2013wlz,Gong:2015iir}.  For the second term appearing in Eq.(\ref{gi_quark_mom}), it has been shown in~\cite{Wilcox:2002zt} that a straight-forward lattice computation of the moments of operators including a spatial coordinate $\vec{x}$ is complicated by periodic boundary conditions on the lattice.
Instead, this contribution has been computed by determining the total angular momentum for the quarks and then subtracting the quark spin contribution to arrive at $L_q$~\cite{Mathur:1999uf,Hagler:2003jd,Bratt:2010jn,Brommel:2007sb,Deka:2013zha}.

On the lattice, the matrix element of $T^{(0i) q,g}$ between two nucleon states can be 
written in terms of three form factors ($T_1, T_2$ and $T_3$) as derived in~\cite{Ji:1996ek},
\bea
\langle p',s' |  T^{\{0i\} q,g} | p,s\rangle
  &=& \frac{1}{2} \bar{u}(p',s') \left[T_1(q^2)(\gamma^0\bar{p}^i 
   +  \gamma^i\bar{p}^0) 
   + \frac{1}{2m}T_2(q^2)\left(\bar{p}^0) (i \sigma^{i\alpha})
   +  \bar{p}^i (i \sigma^{0\alpha})\right) q_{\alpha}\right.\nonumber\\
  &+& \left.\frac{1}{m} T_3(q^2) q^0 q^i\right]^{q,g} u(p,s),
\label{mat_element_1}
\eea
where,\ $p$ and $p'$ are the initial and final momenta of the nucleon,\ respectively,
$\bar{p} = \displaystyle\frac{1}{2}\, (p' + p)$ and $q_\mu = p'_\mu - p_\mu$ is the momentum 
transfer, $m$ is the mass of the nucleon, and $u(p,s)$ is the nucleon spinor. The indices $s'$ and 
$s$ are the initial and final spins, respectively~\cite{Deka:2013zha}.

\noindent
By calculating various polarized and unpolarized three-point functions for Eq.~(\ref{mat_element_1}) at finite $q^2$,
and (\ref{ji_gauge_invariant}),\ and then taking  $q^2 \rightarrow 0$ limit,\ one obtains,
\bea
\label{ang_op_def_split_3}
J^{q,g} &=& \frac{1}{2} \left[T_1(0) + T_2(0)\right]^{q,g},\\
\langle x\rangle^{q,g} &=& T_1(0)^{q,g}.
\label{momentum_fraction}
\eea
where,\ $\langle x\rangle^{q,g} = T_1(0)^{q,g}$ is the first moment of the momentum 
fraction carried by the quarks or glue inside the nucleon.\ 

From Eqs.~(\ref{ang_op_def_split_3})~and~(\ref{momentum_fraction}),\ we write the 
momentum and angular momentum sum rules as,
\bea
  \label{eq:mom_sum_rule}
  T_1 (0)^q + T_1 (0)^g  &=& 1, \\
  \left[ T_1 (0) + T_2 (0) \right]^q 
  + \left[ T_1 (0)  + T_2 (0) \right]^g &=& 1. 
  \label{eq:ang_mom_sum_rule}
\eea
Thus it is clear that to evaluate $J^{q,g}$ (or, $L^{q,g}$), one must compute both the $T_1(0)$ and $T_2(0)$ form factors.  And from Eq.~(\ref{mat_element_1}), these form factors are extracted from the matrix element $\langle p',s' |  T^{\{0i\} q,g} | p,s\rangle$.  In this work, we compute the renormalization and mixing constants associated with these operators at the one-loop level.  As stated in the introduction, this calculation follows similar calculations of the mixing of leading twist operators under the renormalization group.  The essential new piece in this calculation is the introduction of a $T^{\mu\nu}_g$ which is defined from the overlap Dirac operator.  We discuss more details regarding the momentum space operators $T^{\mu\nu}_{q,g}$ in the next section.      

 \section{EM Tensor Operators}\label{operators}
In this section we outline the lattice operators we use for our renormalization calculations based on the discussion in the previous section.  The operators we investigate are similar to leading twist operators in QCD, and can be written compactly,
\bea
\mathcal{O}_{\mu\nu}^q &=& \frac{1}{2}\mathcal{S}\>\sum_f\bar{\psi}_f\left(\gamma_{\mu}  \overleftrightarrow{D}_\nu\right)\psi_f\label{quark_operator}\\
\mathcal{O}^g_{\mu\nu} &=& \frac{1}{2}\mathcal{S}\>\text{tr}_c\>G^{\mu\alpha}G^\nu_\alpha\label{gluon_operator},
\eea
where the symbol $\mathcal{S}$ instructs us to take the symmetrized and traceless piece of the operator, $\overleftrightarrow{D} = 1/2\>(\overrightarrow{D}-\overleftarrow{D})$, and $\text{tr}_c$ is a trace over color indices.  These operators are gauge invariant and we will assume in further discussions that they are symmetrized with respect to all Lorentz indices.  

For the quark operator appearing in~Eq.(\ref{quark_operator}), the covariant derivative is defined from the Wilson action,
\bea
\overrightarrow{D}_\mu \psi(x) &=& \frac{1}{2a}\left(U_{\mu}(x)\psi(x+a \hat{\mu})-U^\dagger_\mu(x-a\hat{\mu})\psi(x-a\hat{\mu})\right),\\
\overleftarrow{D}_\mu \psi(x) &=& \frac{1}{2a}\left(\bar{\psi}(x+a\hat{\mu})U_\mu(x)^\dagger - \bar{\psi}(x-a\hat{\mu})U_\mu(x-a\hat{\mu})\right).
\eea    
where $U_\mu(x) = \exp\left(ig_0a A_{\mu}(x)\right)$ is the link variable at lattice site $x$, with lattice spacing $a$ and coupling $g_0$.  In the quark operator, one can integrate by parts to remove the left-acting derivative in favor of right-acting derivatives only.  An expansion of the link variable in the coupling $g_0$ allows one to write the momentum space vertices necessary for the one-loop renormalization of $\mathcal{O}^q_{\mu\nu}$~\cite{Capitani:1994qn,Capitani:2002mp},
\bea
\mathcal{O}_{\mu\nu}^q = \mathcal{O}_{\mu\nu}^{q,0}+\mathcal{O}_{\mu\nu}^{q,1}+\mathcal{O}_{\mu\nu}^{q,2}+\ldots,
\eea
where,
\bea
\mathcal{O}_{\mu\nu}^{q,0}&=&\frac{1}{2a}\sum_x\left(\bar{\psi}(x)\gamma_\mu\psi(x+a\hat{\nu})-\bar{\psi}(x)\gamma_\mu\psi(x-a\hat{\nu})\right)\label{qo0}\\
\mathcal{O}_{\mu\nu}^{q,1}&=&\frac{ig_0}{2}T^a\sum_x\>\left(\bar{\psi}(x)\gamma_\mu A_\nu^a(x)\psi(x+a\hat{\nu}) + \bar{\psi}(x)\gamma_\mu A_\nu^a(x-a\hat{\nu})\psi(x-a\hat{\nu})\right)\label{qo1}\\
\mathcal{O}_{\mu\nu}^{q,2}&=&-\frac{a g_0^2 }{4}T^aT^b\sum_x\>\left(\bar{\psi}(x)\gamma_\mu A_\nu^a(x)A_\nu^b(x)\psi(x+a\hat{\nu})-\bar{\psi}(x)\gamma_\mu A_\nu^a(x-a\hat{\nu})A_\nu^b(x-a\hat{\nu})\psi(x-a\hat{\nu})\right)\label{qo2}.\nn\\
\eea
In using the notation $\mathcal{O}^{q,i}_{\mu\nu}$, we denote the order in the QCD coupling by the power $i$.  To Fourier transform these operators into momentum space, we define the following Fourier transformations on the quark and gauge fields,
\bea
\psi(x) &=& \int_{-\pi/a}^{\pi/a}\frac{d^4k}{(2\pi)^4}\>e^{ikx}\psi(k),\\
A_\mu(x) &=& \int_{-\pi/a}^{\pi/a}\frac{d^4k}{(2\pi)^4}\>e^{i(x+a\mu/2)k}A_{\mu}(k).
\eea
The complete Feynman rules for each order in the coupling are collected in appendix~\ref{appendixA}.  The Feynman rules for the glue operator involve traces of the overlap Dirac derivative and are thus more cumbersome to compute.  Because of this, we provide more details on our methodology in this section.

Specifically, the field strength tensors which compose the gluon operator $\mathcal{O}^g_{\mu\nu}$ are constructed from the overlap Dirac derivative.  The renormalization constants and mixing coefficients of this operator have not yet been studied in the literature.  Although this operator has been defined from the overlap derivative, one can make contact with the classical field strength tensor.  One can prove that the kernel of the overlap Dirac operator is equivalent to the classical field strength tensor in the continuum limit~\cite{Liu:2007hq},
\bea
\text{tr}_s\>\sigma_{\mu\nu}D_\text{ov}(x,x) = a^2c^T(\rho,r)G_{\mu\nu}(x)+\mathcal{O}(a^3),
\label{trdov}
\eea 
where tr$_s$ denotes a trace over spinor indices, $\sigma_{\mu\nu}=\frac{1}{2i} [\gamma_\mu,\gamma_\nu]$,  $G_{\mu\nu}=g_0\partial_{[\mu}A_{\nu]}-g_0^2[A_\mu,A_\nu]$, and $c^T(\rho)$ is an integration constant given by,
\bea
c^T(\rho,r) &=& \rho\int_{-\pi}^{\pi}\frac{d^4 k}{(2\pi)^4}\frac{2(M c_\mu c_\nu + r s_\mu^2 c_\nu + r s_\nu^2 c_\mu)}{z^{3/2}},\\
z &=& \sum_\mu s_\mu^2 + M^2,\nn\\
\qquad M &=& \rho + r\sum_\mu(c_\mu-1),\nn\\
c_\mu&=&\cos k_\mu,\qquad s_\mu = \sin k_\mu. \nn
\eea
For one-loop calculations, rather than a Taylor expansion in the lattice spacing `$a$' in Eq.(\ref{trdov}), we need an order by order expansion in the coupling constant $g_0$.  For this, we project out the diagonal component of $D_\text{ov}(x,y)$, compute the trace over Lorentz indices, and finally Fourier transform the result in momentum space, order by order in the coupling.

We give here a brief sketch of the procedure used to compute the momentum space Feynman rules of the gluon operator.  The collected results for the lowest order vertices can be found in appendix~\ref{appendixA}.  We follow the methods outlined in~\cite{Fujiwara:2002xh, Liu:2007hq}, and write the diagonal component of the overlap Dirac operator,
\bea
D_{\text{ov}}(x,x) = \sum_y D(x,y)\delta_{xy} = \sum_y\int_{-\pi/a}^{\pi/a}\frac{d^4k}{(2\pi)^4}e^{ikx}D_{\text{ov}}(x,y)e^{-iky}\label{doverlap},
\eea
where we use the following definition for the overlap operator,
\bea
D_{\text{ov}}(x,y) = \frac{\rho}{a}\left(1-X\frac{1}{\sqrt{X^\dagger X}}\right)_{x,y},
\eea
and $X(x,y)$ is the Wilson derivative, which has the discretized form,
\bea
X(x,y) &=& \frac{1}{2a}\sum_\mu\left[\gamma_\mu\left(\delta_{x+\hat{\mu},y}U_\mu(x)-\delta_{x,y+\hat{\mu}}U^\dagger_\mu(y)\right)\right.\nn\\
&&+r\left.\left(2\delta_{x,y}-\delta_{x+\hat{\mu},y}U_\mu(x)-\delta_{x,y+\hat{\mu}}U^\dagger_\mu(y)\right)\right]-\frac{\rho}{a}\delta_{x,y}\label{wilson_d}.
\eea
An expansion, order by order in the coupling constant $g_0$, can be obtained by rewriting the square root term as an integral over a $\sigma$ parameter and Taylor expanding the resulting rational function as a series in the coupling constant~\cite{Ishibashi:1999ik, Capitani:2002mp},
\bea
\frac{1}{\sqrt{X^\dagger X}} = \int_{-\infty}^{\infty}\frac{d \sigma}{\pi}\frac{1}{\sigma^2+X^\dagger X}\label{ov_den}.
\eea
The product $X^\dagger X$ can be expanded (in powers of $g$) order by order, we introduce the following shorthand,
\bea
X^\dagger X=\sum_i g_0^i(X^\dagger X)_i \equiv \sum_i g_0^i\> \Sigma_i,
\eea
to the lowest orders we have, then,
\bea
\Sigma_0 &=& X^\dagger_0 X_0\\
\Sigma_1 &=& X^\dagger_0 X_1+X^\dagger_1X_0\\
\Sigma_2 &=& X^\dagger_0 X_2+X^\dagger_2X_0+X^\dagger_1X_1,
\eea 
where the subscripted $\Sigma_i$ and $X_i$ indicate at which order in the QCD coupling the various $\Sigma$ factors have been expanded .  The expressions for the various $X_i$ operators in momentum space can be found in append~\ref{appendixA}.  With these results, we Taylor expand Eq.(\ref{ov_den}) order by order in the coupling $g_0$.
For example, the zeroth, first and second order expansions are,
\bea
\left(\frac{1}{\sigma^2+X^\dagger X}\right)_0&=&\frac{1}{\sigma^2+\Sigma_0}\\
\left(\frac{1}{\sigma^2+X^\dagger X}\right)_1&=&-\frac{1}{\sigma^2+\Sigma_0}\Sigma_1\frac{1}{\sigma^2+\Sigma_0}\nn\\
\left(\frac{1}{\sigma^2+X^\dagger X}\right)_2&=&\frac{1}{\sigma^2+\Sigma_0}\Sigma_1\frac{1}{\sigma^2+\Sigma_0}\Sigma_1\frac{1}{\sigma^2+\Sigma_0}-\frac{1}{\sigma^2+\Sigma_0}\Sigma_2\frac{1}{\sigma^2+\Sigma_0}.\nn\\
\eea
Examining the form of the $\Sigma_i$, we can see that the zeroth order expansion of $D_\text{ov}$ will vanish when traced over $\sigma_{\mu\nu}$,
\bea
\left(\text{tr}\> \sigma_{\mu\nu}D_{\text{ov}}\right)^0 \equiv G_{\mu\nu}^0(x,x)  = \frac{\rho}{\pi a}\> \text{tr}\> \sigma_{\mu\nu} \int_{-\infty}^\infty d\sigma \sum_y \int_k e^{ikx}X_0\>\frac{1}{\sigma^2 + \Sigma_0}\>e^{-iky}.
\eea  
The Dirac structure of $X_0$ is $X_0 = A \gamma_\mu + B$ where both $A$ and $B$ are Lorentz scalars, and $\Sigma_0 = X^\dagger_0 X_0$ is also a Lorentz scalar, see appendix~\ref{appendixA} for details.  Thus, when traced over $\sigma_{\mu\nu}$, this expression vanishes.  

The various products $X/\sqrt{X^\dagger X}$ expanded to the next three lowest orders in the coupling $g$ are listed below.  The third order expansion is necessary to calculate tadpole contributions to the renormalization constant $Z_{G\rightarrow G}$ which contains a fourth-order vertex.  After the taylor expansion and noting that $\Sigma_0$ is a commuting object we find for the first three orders of the expansion of $\text{tr}\>\sigma_{\mu\nu}D_{\text{ov}}(x,x)\equiv G_{\mu\nu}(x,x)$ in Eq.(\ref{trdov}),
\begin{small}
\bea
\left(\text{tr}\> \sigma_{\mu\nu}D_{\text{ov}}\right)^1& \equiv& G^1_{\mu\nu}(x,x)  = g_0\frac{\rho}{\pi a}\>\text{tr}\>\sigma_{\mu\nu}\int_{-\infty}^\infty d\sigma\sum_y \int_k e^{ikx}\>\hat{\Pi}\left(\sigma^2 X_1 - X_0X_1^\dagger X_0\right)\hat{\Pi}\>e^{-iky}\label{oo1}\\
\nn\\
\nn\\
\left(\text{tr}\> \sigma_{\mu\nu}D_{\text{ov}}\right)^2 &\equiv& G^2_{\mu\nu}(x,x)  = g_0^2\frac{\rho}{\pi a}\>\text{tr}\>\sigma_{\mu\nu}\int_{-\infty}^\infty d\sigma\sum_y \int_k e^{ikx}\left( \hat{\Pi} \left\{\sigma^2 X_2 - X_0 X_2^\dagger X_0\right\}\hat{\Pi}\right.\nn\\
 &&\left.-\hat{\Pi}\left\{\sigma^2(X_1 X_0^\dagger X_1 + X_1 X_1^\dagger X_0 + X_0 X_1^\dagger X_1)- X_0 X_1^\dagger X_0 X_1^\dagger X_0\right\}\hat{\Pi}^2\right)e^{-iky}\label{oo2}\\
 \nn\\
\left(\text{tr}\> \sigma_{\mu\nu}D_{\text{ov}}\right)^3 &\equiv& G^3_{\mu\nu}(x,x) = g_0^3\frac{\rho}{\pi a} \>\text{tr}\>\sigma_{\mu\nu}\int_{-\infty}^{\infty}d \sigma\sum_{y}\int_k e^{ikx}\left(\hat{\Pi}\left\{\sigma^2 X_3 -\mathcal{A}\right\}\hat{\Pi} + \hat{\Pi}^2 \left\{\sigma^2 \mathcal{B}+\mathcal{C}\right\}\hat{\Pi}^2\right)e^{-iky}\label{o3}\nn\\,
\eea
\end{small}
where again the power $i$ in $G^i$ denotes the order in the QCD coupling.  We have made use of the shorthand, $\int_k \equiv \int \text{d}^4k/(2\pi)^4$ and $\mathcal{A}, \mathcal{B}$ and $\mathcal{C}$ in $G_{\mu\nu}^3$ are lengthy expressions involving products of $X_i$, and $\hat{\Pi}=\frac{1}{\sigma^2 + \Sigma_0}$.  The exact forms for $\mathcal{A}, \mathcal{B}$ and $\mathcal{C}$ can be found in appendix \ref{appendixA}. 
Before we Fourier transform each order in the coupling $g$, we compute the action of the various $X_i$ derivatives on $e^{-iky}$ as shown in Eq.(\ref{doverlap}), we have, using Eq.(\ref{wilson_d}),
\bea
X e^{-i ky} f(x) &=& e^{-iky} \left\{ \sum_\mu\gamma_\mu \left(\tilde{Q}_\mu - \frac{i}{a}s_\mu\right)-r_w \sum_\mu\left(-\frac{1}{a}(1-c_\mu)+\tilde{R}_\mu\right)-\frac{\rho}{a}\right\}f(x)\label{xoperator},
\eea
where, 
\bea
\tilde{Q}_\mu &=& \frac{1}{2}\left(e^{-ik_\mu} \nabla_\mu + e^{ik_\mu}\nabla^*_\mu\right)\\
 \tilde{R}_\mu &=& \frac{1}{2}\left(e^{-ik_\mu}\nabla_\mu - e^{ik_\mu}\nabla^*_\mu\right)
\eea
and,
\bea
\nabla_\mu\psi(x) &=& \frac{1}{a}\left(U_\mu(x) \psi(x+a\hat{\mu}) -\psi(x)\right)\\
\nabla^*_\mu\psi(x) &=& \frac{1}{a}\left(\psi(x)-U^{\dagger}_\mu(x-a\hat{\mu})\psi(x-a\hat{\mu})\right).
\eea
Eqs.(\ref{qo0}-\ref{qo2}) as well as Eqs.(\ref{oo1}-\ref{o3}) are the main results for this section.  For the glue operator, what remains is to compute, order by order the products acting on the unit vector $1_n$,
\bea
P_i = X_0 X_i^\dagger X_0 \frac{1}{\sigma^2+X_0^\dagger X_0}\hat{1}
\eea
which appear in Eqs.(\ref{oo1},\ref{oo2}), Fourier transform all gauge fields to momentum space, and finally compute the trace over the Dirac indices.  These details are somewhat lengthy and are relegated to appendix~\ref{appendixA} for the interested reader.  We close this section by remarking that once these calculations are performed, we can construct the full momentum space gluon operator $G_{\mu\alpha}G^{\alpha}_\nu$ order by order in the coupling by using our results for the field strength tensor.  We can write this expansion schematically,
\bea
\mathcal{O}_{\mu\nu}^g = \left(G_{\mu\alpha}^0 + g_0G_{\mu\alpha}^1+g_0^2 G_{\mu\alpha}^2+\ldots\right)\left(G^{0,\alpha}_\nu+g_0G^{1,\alpha}_\nu+g_0^2G^{2,\alpha}_\nu +\ldots\right). 
\eea  
We note that the trace over $\sigma_{\mu\nu}$ in Eq.(\ref{trdov}) causes all terms involving $G_{\mu\alpha}^0$ to vanish.  At lowest order, then, we have Feynman rules for two and three external gauge fields respectively, 
\bea
\mathcal{O}^{g,2}_{\mu\nu} &=& g_0^2 G_{\mu\alpha}^1 \> G_{\alpha}^{\nu,1} \\
\nn\\
\mathcal{O}^{g,3}_{\mu\nu} &=& g_0^3 \left(G_{\mu\alpha}^1 \> G_{\alpha}^{\nu,2} +  G_{\mu\alpha}^2 \> G_{\alpha}^{\nu,1}\right),
\eea 
where a symmetrization over Lorentz indices and a trace over color indices has been suppressed. 
 \section{Renormalization}\label{renormalization}
\begin{figure}
\begin{center}
\includegraphics[scale=0.9]{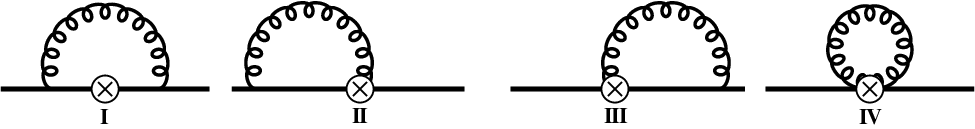}
\caption{Feynman diagrams for the calculation of $Z_{Q\to Q}$.  The circle represents an insertion of the twist-2 operator.}
\label{zqq}
\end{center}
\end{figure}

\begin{figure}
\begin{center}
\includegraphics[scale=0.75]{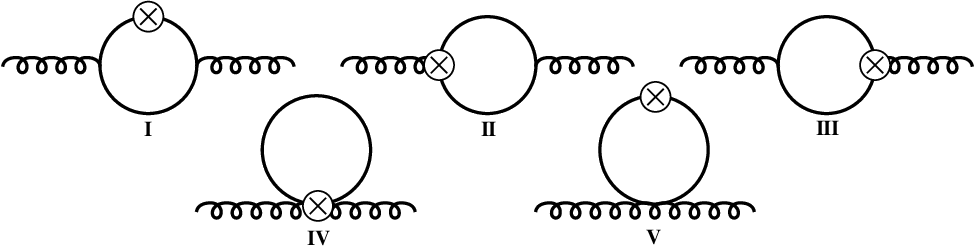}
\caption{Feynman diagrams for the calculation of $Z_{Q\rightarrow G}$.  The circle represents an insertion of the quark angular momentum operator.}
\label{zgq}
\end{center}
\end{figure}

\begin{figure}
\begin{center}
\includegraphics[scale=0.65]{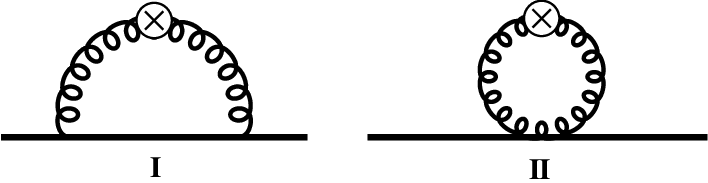}
\caption{Feynman diagrams for the calculation of $Z_{G\to Q}$.  The circle represents an insertion of the glue EM tensor operator defined from the overlap Dirac derivative.}
\label{zqg}
\end{center}
\end{figure}

\begin{figure}
\begin{center}
\includegraphics[scale=0.7]{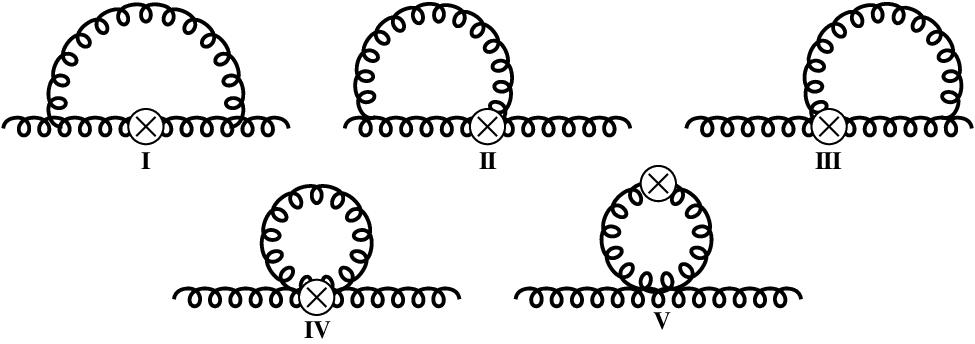}
\caption{Feynman diagrams for the calculation of $Z_{G\rightarrow G}$.  The circle represents an insertion of the gluon angular momentum operator defined from the overlap Dirac derivative.}
\label{zgg}
\end{center}
\end{figure}

 In this section we detail the renormalization conditions used in our calculations.  We remark that since we are calculating the one-loop corrections to flavor-singlet operators, the gluon operator is allowed to mix with the quark operator beyond tree level.  This renormalization and mixing arise from diagrams like those shown in Figs.[\ref{zqq},\ref{zgg}] and Figs.[\ref{zgq},\ref{zqg}] respectively.  Due to these diagrams, the renormalization constants $Z$ are in fact matrices $Z_{ij}$, and we can organize our calculation in a $2\times 2$ matrix
 \bea
 \mathcal{O}_{i}^r = \sum_j Z_{ij} \mathcal{O}^b_j,
 \eea
 where the superscript $b$ denotes a bare operator and $r$ on the LHS denotes the renormalized operator.  The indices $i$, $j$ run over the operator basis.  As in the continuum, we denote the renormalization factors for the massless fermion wave function and strong coupling constant as $Z_\psi$ and $Z_g$ respectively,
 \bea
 \psi_b = \sqrt{Z_\psi} \psi_r,\qquad A_b = \sqrt{Z_A} A_r, \qquad  g_b = Z_g g_r.
 \eea  
For both the bare wave function and the bare coupling we have used the notation $\psi_0$ and $g_0$ respectively.  These renormalization constants can be expanded around unity,
 \bea
 Z_\psi = 1 + \delta Z_\psi, \qquad Z_A = 1 + \delta Z_A, \qquad Z_g = 1+\delta Z_g,
 \eea
 where $\delta Z_\psi$ and $\delta Z_g$ denote the contributions from higher order diagrams.  Similarly, the $Z_{ij}$ renormalization constants can be expanded around unity,
 \bea
\> Z_{Q\rightarrow Q} = 1 + \delta Z_{Q\rightarrow Q}, \qquad Z_{Q\rightarrow G} = 1 + \delta Z_{Q\rightarrow G},\\\
  Z_{G\rightarrow Q} = 1 + \delta Z_{G\rightarrow Q}, \qquad Z_{G\rightarrow G} = 1 + \delta Z_{G\rightarrow G}.
 \eea
 %
 %

\newpage
 \subsection{Quark EM Tensor}
 The bare quark angular momentum operator has the schematic form,
 \bea
 \mathcal{O}^b_Q = g_b\bar{\psi}_b \psi_b,
 \eea
where the Lorentz structure and various derivative terms have been omitted.  Throughout the one-loop calculations, the renormalization constants $Z_{ij}$ appearing in the previous section are fixed by a set of renormalization conditions on the quark and gluon matrix elements.  For the quark operator, the renormalized and bare quark matrix elements are related as,
 \bea\label{qrc}
 \langle \bar{\psi}_r | \mathcal{O}^r_Q(\mu) | \psi_r \rangle |_{p^2=\mu^2} &=& Z_{Q\rightarrow Q}(a\mu,g_b)\> Z_\psi^{-1}(a\mu,g_b)\>\langle \bar{\psi}_b | \mathcal{O}_Q^b(a) | \psi_b \rangle^{\text{1-loop}} \nn\\
 &&+Z_{Q\rightarrow G}(a\mu, g_b) \>\langle A_b,\lambda| \mathcal{O}^b_{Q}(a)| A_b,\lambda\rangle^{\text{1-loop}}\nn\\
 &\equiv& \langle \bar{\psi}_b | \mathcal{O}^b_Q(a) |\psi_b \rangle^{\text{tree}},
 \eea
 where $\lambda$ is a polarization index for the external gauge field.  The tree level matrix element $\langle \bar{\psi}_b | \mathcal{O}^b_Q(a) |\psi_b \rangle^{\text{tree}}$, is defined by,
 \bea\label{quark-tree}
\langle \bar{\psi} | \mathcal{O}^Q_{\mu\nu} | \psi \rangle_{\text{tree}} &=& \frac{i}{2} \left( \gamma_\mu p_\nu + \gamma_\nu p_\mu\right).
\eea
 With this renormalization condition, the renormalization constants $Z_{Q\rightarrow Q}$ and $Z_{Q\rightarrow G}$ are fixed by computing the diagrams shown in Fig.[\ref{zqq}] and Fig.[\ref{zgq}] respectively.  While, the $Z_\psi$ is fixed from wave function renormalization of the quark field.  In Eq.[\ref{qrc}], we have made use of the fact that the tree-level matrix elements,
 \bea
 \langle \bar{\psi}_b | \mathcal{O}^b_G(a) |\psi_b \rangle^{\text{tree}}, \>\>\>\>  \langle A_b | \mathcal{O}^b_Q(a) |A_b \rangle^{\text{tree}}.
 \eea   
 both vanish.
 %
 %
 
%
\newpage
 \subsection{Glue EM Tensor}
 The bare gluon operator has the schematic form,
 \bea
 \mathcal{O}^b_G = G_0 G_0.
 \eea
 The renormalized and bare gluon operators are then related,
  \bea
 \langle A_r,\lambda | \mathcal{O}^r_G(\mu) | A_r,\lambda \rangle |_{p^2=\mu^2} &=& Z_{G\rightarrow G}(a\mu,g_b)\> Z_A^{-1}(a\mu,g_b)\>\langle A_b,\lambda | \mathcal{O}_G^b(a) | A_b,\lambda \rangle^{\text{1-loop}} \nn\\
 &&+Z_{G\rightarrow Q}(a\mu, g_b) \>\langle \bar{\psi}_b| \mathcal{O}^b_{G}(a)| \psi_b \rangle^{\text{1-loop}}\nn\\
 &\equiv& \langle A_b,\lambda | \mathcal{O}^b_G(a) |A_b,\lambda \rangle^{\text{tree}}.
 \eea

 As with the quark operator, the renormalization constant $Z_{G\rightarrow Q}$ is an off-diagonal mixing term fixed by the diagrams shown in Fig.[\ref{zqg}], and the $Z_{G\rightarrow G}$ renormalization constant is computed from the diagrams shown in Fig.[\ref{zgg}].  Again, the matrix element $\langle A_b,\lambda | \mathcal{O}_Q | A_b,\lambda \rangle $ vanishes at tree-level, but is non-zero at one-loop order.  Here, the tree-level matrix element, $\langle A_b,\rho | \mathcal{O}^b_G(a) |A_b,\tau \rangle^{\text{tree}}$, is defined by,
\bea\label{glue-tree}
&&-2p^{\mu}p^{\nu}g^{\rho\tau} + p^{\mu}p^{\rho}g^{\nu\tau}-p^{\mu}p^{\nu}g^{\rho\tau}-p^2g^{\rho\mu}g^{\nu\tau}+p^{\tau}p^{\nu}g^{\rho\mu}+p^{\nu}p^{\rho}g^{\mu\tau}-p^2g^{\rho\nu}g^{\mu\tau}+p^{\tau}p^{\mu}g^{\rho\nu}\nonumber\\
&&\qquad-g^{\mu\nu}(p^{\tau}p^{\rho}-p^2g^{\tau\rho})\ .
 \eea
.  We point out that in the final stages of all one-loop calculations we encounter complicated expressions depending on the external momentum and possibly Dirac gamma matrices.  These expressions must be grouped into gauge-invariant terms representing the tree-level matrix elements of the quark and gluon EM tensor operators defined in Eqs.(\ref{quark-tree}, \ref{glue-tree}) before we can extract the correct renormalization constants.  

We can simplify the procedure greatly by exploiting our freedom to choose 
\bea\label{eq:condition}
\mu \ne \nu, \sigma=\tau\ne \mu,
\eea
in all calculations~\cite{Capitani:1994qn}, and thus setting all $\delta_{\mu\nu}$ terms to zero. This has the benefit of avoiding any mixing into lower dimensional operators which have the same symmetries under the hypercubic group $H(4)$ as our quark and gluon angular momentum operators. Note that with this condition, the renormalization we obtained in this work cannot be used for the operators $\mathcal{O}_{G,\ \mu\nu}$ with $\mu=\nu$, since they belong to the different irreducible representation of the hypercubic group~\cite{Caracciolo:1991cp}. But it is enough for the proton spin decomposition in Ref.~\cite{Deka:2013zha} since only the off-diagonal piece of $\mathcal{O}_{G,\ \mu\nu}$ were used. See~\cite{yang:2016lpt} for detailed discussion and updates on this point.

We close this section by listing schematic forms for all renormalization constants.  The numerical results for the finite contributions of those $Z$-factors involving the glue operator are found in Tabs.[\ref{glue_results_1}, \ref{glue_results_2}], and those involving the quark operator can be found in Tab.[\ref{quark_results_1}], our results for the various $Z_A$ and $Z_\psi$ are tabulated in appendix \ref{appendixB}.  Schematically, we write,
\bea
Z_{Q\rightarrow Q} &=& 1+\frac{g_0^2}{16\pi^2}C_F \left(\frac{8}{3}\log(a^2 p^2) + F_{Q\rightarrow Q}(r_w)\right) \\
Z_{G\rightarrow Q} &=& -\frac{g_0^2}{16\pi^2}C_F \left(\frac{8}{3}\log(a^2 p^2) + F_{G\rightarrow Q}(\rho,r_w)\right) \\
Z_{Q\rightarrow G} &=& -\frac{g_0^2}{16\pi^2}N_F \left(\frac{2}{3}\log(a^2 p^2) + F_{Q\rightarrow G}(r_w)\right) \\
Z_{G\rightarrow G} &=& 1+\frac{g_0^2}{16\pi^2}\left(\frac{2}{3}N_F\log(a^2 p^2) + N_F F_{G\rightarrow G}(r_w) + N_c B_{G\rightarrow G}(\rho,r_w)\right). 
\eea
 \section{Methodology}\label{methodology}
In this section we outline the methods used to compute the one-loop mixing coefficients outlined in the previous section.  At one-loop order, and after suitable simplification of all Dirac and color matrices, all lattice integrations encountered in this mixing calculation can be expressed in the schematic form,
\bea
I(p) = \int_{-\pi/a}^{\pi/a}\frac{d^4 k}{(2\pi)^4} \frac{\mathcal{N}(k,p)}{\mathcal{D}(k,p)} \equiv \int_k \mathcal{I}(k,p)\label{integral},
\eea
where we have suppressed both the color and Lorentz indices.  The integrand $\mathcal{I}$ is, in general, a complicated rational function of both $k$ and $p$ involving many $\sin$ and $\cos$ terms.  A direct integration of such a function is typically impractical.  Instead one can still achieve a high accuracy result by `splitting' the integrand in the following way,
\bea
I = J + (I - J),
\eea
where $J$ is given by a Taylor expansion in the external momentum $p$,
\bea
J = \sum_{n=0}^N \frac{p_{\mu_1}\ldots p_{\mu_n}}{n!}\left\{\frac{\partial^n}{\partial_{\mu_1}\ldots \partial_{\mu_n}}\>\mathcal{I}(k,p)\right\}_{p\to 0}.
\eea
The order $N$ in this expansion is set by the degree of divergence $\mathcal{I}(k,p)$.  With this result, using the power counting theorem of Reisz, we can compute the difference,
\bea
\lim_{a\to 0}(I-J),
\eea 
in the continuum limit.  For these calculations, the one-loop calculations in the continuum are straightforward.  We point out, however, that the Taylor expansion and artificial splitting of the integrand introduce an infrared divergence at intermediate stages of the calculations.  We have chosen to regulate this divergence using dimensional regularization in $d = 4-2\epsilon$ dimensions.  Thus, we expect both $J$ and $(I-J)$ to exhibit poles in epsilon which must cancel to give a finite result for $I$ at the end of the calculations.  

The Taylor expansion has reduced $J$ to an integral over the loop momentum $k$ only, greatly simplifying its calculation.  However we must still isolate all pole terms and separate them before passing $J$ to any numerical integration routine.  To do so, our scripts reduce $J$ to the following schematic form,
\bea
J = \int \frac{d^dk}{(2\pi)^d}\frac{\mathcal{N}(k)}{D_b^{n_b}(k)\> D_q^{n_q}(k)},
\eea  
where the exact form of the numerator is not important, only that it depends only on $k$, and $D_b$ is the inverse gluon propagator and $D_q$ is a generic inverse quark propagator.  We can isolate any divergent terms in this integrand by writing,
\bea
\frac{1}{D_q(k)} = \frac{1}{D_b(k)} + \left(\frac{1}{D_q(k)}-\frac{1}{D_b(k)}\right)
\eea
The degree of divergence of $(1/D_q - 1/D_b)$ is reduced by one.  By iteratively applying this kind of splitting and separating out integrals involving only powers of $\frac{1}{D_b}$, all pole terms in $J$ can be isolated.  In the end, any $J$ integral involving arbitrary powers of quark and gluon propagators can be expressed as a sum,
\bea
J = \underbrace{\sum \frac{\mathcal{N}(k)}{D_b^{n_b}(k)}}_{\text{divergent}} + \underbrace{\sum\frac{\mathcal{N}'(k)}{D_b^{n'_b}D_q^{n'_q}}}_{\text{finite}}.
\eea
The divergent pieces of this sum can be computed to arbitrary accuracy by using the results in~\cite{Caracciolo:1991cp}.  The remaining finite piece is computed to 9-digit accuracy using the Clenshaw-Curtis algorithm in Mathematica.  At the end of the calculation, all $J$-type integrals can be expressed in a schematic form,
\bea
J = \frac{g_0^2}{16\pi^2}\left(\frac{N}{\epsilon} + F\right)
\eea
where $N$ and $F$ are numerical constants and any Lorentz or color indices have been suppressed. 

As discussed in previous perturbative calculations on the lattice~\cite{Capitani:1994qn, Capitani:2002mp}, a major obstacle in performing these calculations analytically is that gauge field theories regulated by a lattice spacing respect hypercubic symmetries rather than the more restrictive Lorentz symmetries.  This is problematic when trying to apply pre-built packages such as {\tt FORM} to simplify intermediate expressions.  For example, many terms common to lattice perturbation theory, such as $\sum_\mu \gamma_\mu \sin k_\mu \cos p_\mu$ are not properly handled by the existing index contraction methods designed for continuum calculations.  Because of this, we have written several separate scripts in {\tt python} to aid in simplifying intermediate expressions involving products of Dirac matrices in $d$-dimensions before passing the results to our integration routines.  

The programs thus arrive at the final integrated result for $I(p)$ shown in Eq.(\ref{integral}) by first Taylor expanding the momentum space vertices in the external momentum to the desired order.  At this stage all $d$-dimensional gamma algebra is carried out in {\tt FORM} with the aid of several {\tt python} scripts.  Once this has completed, the lattice integral of interest has been expressed as a sum of integrands of the following form,
\bea
I(\mu_1,\ldots , \mu_n) = \int d^4k \> f\left(\sum_\lambda\sin^2 k_\lambda\right)\prod_{i}\sin k_{\mu_i},
\eea 
where $f$ denotes some even function of $\sin$ and odd powers of $\sin$ have integrated to zero by symmetry.  As outlined in~\cite{Capitani:1994qn}, it is advantageous to simplify these products of $\sin$ functions using hypercubic ($H(4)$) symmetries.  We have written {\tt FORM} routines to carry this out automatically.  The details of this stage of the calculation are the same as in~\cite{Capitani:1994qn} and can be found there.   Once these symmetry relations are applied, the integrands are ready to be reduced to their divergent and finite parts.  We have automated this procedure as well with additional {\tt python} code which follows the `splitting' methods described previously in this section.  Finally when all finite pieces have been isolated from the divergent parts, all divergent pieces are simplified analytically using the reduction methods described in~\cite{Caracciolo:1991cp}, and all finite pieces are passed to Mathematica to be integrated, which then collects the final, simplified result.  A crucial check on this method is that the continuum integration $(I-J)$ produces an $\epsilon$-pole which cancels the pole computed in $J$, we show in the next section that this is indeed the case for all the calculations performed.  

We close this section with a brief comment regarding the gauge dependence of these results.  In all one-loop calculations, we have set the gauge parameter $\alpha$ appearing in the gluon propagator (see appendix \ref{appendixD}) to unity, corresponding to the Feynman gauge.  All calculations in this work are in the Feynman gauge and the self consistent check for the general gauge will be addressed in the upcoming work~\cite{yang:2016lpt}.
 \section{Results}\label{results}
 In this section, we report the results for the $Z_{Q\rightarrow Q}$, $Z_{Q\rightarrow G}$, $Z_{G\rightarrow Q}$ and $Z_{G\rightarrow G}$, 
 \begin{small}
\bea\label{eq:renorm}
Z_{Q\rightarrow Q}^{\text{Latt.}} &=& 1+\frac{g^2_{0,\text{Latt.}}}{16\pi^2}C_F \left(\frac{8}{3}\log(a^2 p^2) + F_{Q\rightarrow Q}(r_w)\right) \\
Z_{G\rightarrow Q}^{\text{Latt.}} &=& -\frac{g^2_{0,\text{Latt.}}}{16\pi^2}C_F \left(\frac{8}{3}\log(a^2 p^2) + F_{G\rightarrow Q}(\rho,r_w)\right) \\
Z_{Q\rightarrow G}^{\text{Latt.}} &=& -\frac{g^2_{0,\text{Latt.}}}{16\pi^2}N_f \left(\frac{2}{3}\log(a^2 p^2) + F_{Q\rightarrow G}(r_w)\right) \\
Z_{G\rightarrow G}^{\text{Latt.}} &=& 1+\frac{g^2_{0,\text{Latt.}}}{16\pi^2}\left(\frac{2}{3}N_f\log(a^2 p^2) + N_f F_{G\rightarrow G}(r_w) + N_c B_{G\rightarrow G}(\rho,r_w)\right).
\eea
\end{small}
where $N_c$ and $N_f$ are the number of colors and flavors respectively.  The results of the finite pieces $F$ and $B_G$ are summarized in tables \ref{quark_results_1}, \ref{glue_results_1}, \ref{glue_results_2}. For completeness the expressions for $Z_g$ and  $Z_\psi$ needed to compute the final values for the renormalization constants in Eq.[\ref{eq:renorm}] are listed in appendix~\ref{appendixB}.
  
For the case of $Z_{Q\rightarrow Q}$ and $Z_{G\rightarrow Q}$, the related diagrams do not involve the glue EM tensor operator, see Figs.[\ref{zqq}, \ref{zgq}], and have been calculated in~\cite{Capitani:1994qn}.  Our results of $F_{Q\rightarrow Q}$ have good agreement with those in ~\cite{Capitani:1994qn}, but the results of $F_{Q\rightarrow G}$ are different. Due to the mixing with the glue equation of motion term, the finite piece under RI-MOM scheme in the continuum depends on the momentum on the external legs as $-\frac{4}{9}-\frac{2}{3}p^{\mu}p^{\nu}p^{\rho}p^{\tau}$ where $p$ is the momentum of the external legs and $\mu/\nu$ and $\rho/\tau$ are  the indices of the operator and external legs respectively. We confirm that our results have the same external momentum dependence as that in the continuum and then the final renormalization constant under $\overline{MS}$ scheme is a constant only related to the UV regulator. We take $p^{\rho/\tau}=0$ in the rest part of this work to simplify the expression. 

The results of those diagrams containing the glue EM tensor operator (for the case of $Z_{G\rightarrow G}$ and $Z_{G\rightarrow Q}$) are shown in Figs.[\ref{zgg}, \ref{zqg}].  This operator has been constructed from the overlap Dirac derivative and its renormalization has not yet been studied in the literature.  Our results depend on several parameters, specifically $0 < r_w \le 1$ and $0 < \rho < 2r_w$ from Eq.(\ref{wilson_d}).  We quote the results for several values of $\rho$ and allow $r_w$ to vary from 0.2-1 in increments of 0.2.  We emphasize that all color factors have been divided out of these results, along with an overall factor of $1/(16\pi^2)$ and as well as the tree-level expression for the operator of interest.  %

\begin{table}[H]
\centering
\begin{ruledtabular}
\begin{tabular}{rccc}
$Z_{Q\rightarrow Q}^{\text{Latt.}}$ (Fig.(\ref{zqq})) & $r_w$ & $F_{Q\rightarrow Q}(r_w)$& \\
 \hline
  &0.2 & $-7.5170$ &\\
  &0.4 & $-6.3690$ &\\ 
   &0.6 &$-5.1610$ &\\ 
     &0.8 & $-4.0900$&\\ 
     &1.0 &  $-3.1649$&\\     
    \hline 
$Z_{Q\rightarrow G}^{\text{Latt.}}$ (Fig.(\ref{zgq})) & $r_w$& $F_{Q\rightarrow G}(r_w)$& \\
 \hline
& 0.2 & $\ 0.5542$ &\\
 & 0.4 & -0.0960 &\\ 
  & 0.6 & -0.1111&\\ 
    & 0.8 & $\ 0.0322$ &\\ 
    & 1.0 & $\ 0.2078$ &\\ 
\end{tabular}
\caption{ Results for the $Z_{Q \rightarrow Q}$ and $Z_{Q\rightarrow G}$ mixing calculation.  These results have been computed previously in~\cite{Capitani:1994qn}, we have found agreement for $F_{Q \rightarrow Q}$. However our $F_{Q \rightarrow G}$ are different from those in Ref.~\cite{Capitani:1994qn}.}
\label{quark_results_1}
\end{ruledtabular}
\end{table}
     
\begin{table}[H]
\centering
\begin{ruledtabular}
\begin{tabular}{rcccc}
$Z_{G\rightarrow Q}^{\text{Latt.}}$ (Fig.(\ref{zqg})) & $r_w$&$F_{G\rightarrow Q}(\rho=1,r_w)$& -- &\\
\hline
  & 0.2 &  $4.06025$ &&\\
 & 0.4 & $3.39754$ &&\\
  & 0.6 &  $2.88773$ &&\\
    & 0.8 &  $2.38546$ &&\\
    & 1.0 &   $1.90172$ &  $$\\    
    \hline
    $Z_{G\rightarrow G}^{\text{Latt.}}$ (Fig.(\ref{zgg}))& $r_w$ & $F_{G\rightarrow G}(r_w)$& $B_{G\rightarrow G}(\rho=1,r_w)$&\\
 \hline
  & 0.2 & $1.22383$   &$-1.18448$   &\\
 & 0.4 &  $1.36776$  &$-1.23117$   &\\
  & 0.6 & $1.60728$  &$-1.28174$   &\\
    & 0.8 &  $1.97383$   &$-1.33272$  &\\
        & 1.0 & $2.16850 $ &  $-1.38353$ & \\     
\end{tabular}
\caption{Results for the mixing constants $Z_{G\rightarrow G}$ and $Z_{G\rightarrow Q}$.  In this table, we have chosen $\rho = 1$ and have listed results for several values of the Wilson $r_w$ parameter.  }
\label{glue_results_1}
\end{ruledtabular}
\end{table}

\begin{table}[H]
\centering
\begin{ruledtabular}
\begin{tabular}{rcccc}
$Z_{G\rightarrow Q}^{\text{Latt.}}$ (Fig.(\ref{zqg})) & $r_w$&$F_{G\rightarrow Q}(\rho=1.368,r_w)$& -- &\\
\hline
  & 0.2 &  $5.28282$ &&\\
 & 0.4 & $5.10614$ &&\\
  & 0.6 &  $4.96733$ &&\\
    & 0.8 &  $4.86544$ &&\\
    & 1.0 &   $4.82048$ &  $$\\    
    \hline
    $Z_{G\rightarrow G}^{\text{Latt.}}$ (Fig.(\ref{zgg}))& $r_w$ & $F_{G\rightarrow G}(r_w)$& $B_{G\rightarrow G}(\rho=1.368,r_w)$&\\
 \hline
  & 0.2 & $1.22383$   &$-0.104783 $   &\\
 & 0.4 &  $1.36776$  &$-0.105484 $   &\\
  & 0.6 & $1.60728$  &$-0.106373 $   &\\
    & 0.8 &  $1.97383$   &$-0.107885$  &\\
        & 1.0 & $2.16850 $ &  $-0.108396$ & \\     
\end{tabular}
\caption{Results for the mixing constants $Z_{G\rightarrow G}$ and $Z_{G\rightarrow Q}$.  Here, we have chosen $\rho = 1.368$, and have listed results for several values of the Wilson $r_w$ parameter.}
\label{glue_results_2}
\end{ruledtabular}
\end{table}

Here, we report the continuum $\overline{\text{MS}}$ finite contributions necessary to match our lattice renormalization mixing constants to the continuum $\overline{\text{MS}}$ scheme with the mathematica package Package-X~\cite{Patel:2015tea},  
\bea
Z^{\overline{\text{MS}}}_{Q\rightarrow Q} &=& 1+\frac{g_{0,\overline{\text{MS}}}^2}{16\pi^2}C_F \left(\frac{8}{3}\log( p^2/\mu^2) - \frac{40}{9}\right) \\
Z^{\overline{\text{MS}}}_{G\rightarrow Q} &=& -\frac{g_{0,\overline{\text{MS}}}^2}{16\pi^2}C_F \left(\frac{8}{3}\log(p^2/\mu^2) -\frac{22}{9}\right) \\
Z^{\overline{\text{MS}}}_{Q\rightarrow G} &=& -\frac{g_{0,\overline{\text{MS}}}^2}{16\pi^2}N_f \left(\frac{2}{3}\log(p^2/\mu^2) -\frac{4}{9}\right) \\ 
Z^{\overline{\text{MS}}}_{G\rightarrow G} &=& 1+\frac{g_{0,\overline{\text{MS}}}^2}{16\pi^2}\left(\frac{2}{3}N_f\log( p^2/\mu^2) - N_f \frac{10}{9} + N_c \frac{4}{3}\right). 
\eea
So the final matching factors are computed from the condition in Ref.~\cite{Capitani:2002mp} as,
\bea
Z_{i\rightarrow j}^{ \text{Latt.}\rightarrow\overline{\text{MS}}}(a \mu, g_0) &=& \sum_k Z_{i\rightarrow k}^{\overline{\text{MS}}}\left(p^2/\mu^2,g_{0,\overline{\text{MS}}}\right)\>Z_{k\rightarrow j}^{-1,\text{Latt.}}\left(p^2 a^2,g_{0,\text{Latt.}}\right)\\
&=& \left(\delta_{i\rightarrow j} + \frac{g_{0,\text{Latt.}}^2}{16\pi^2}(\gamma_{i\rightarrow j} \log a^2\mu^2 + F_{i\rightarrow j}^{\overline{\text{MS}}}-F_{i\rightarrow j}^{\text{Latt.}})\right).
\eea
In the above matching condition, we have chosen to take the coupling to be the lattice coupling, as is conventional.  The difference between the lattice and continuum couplings only appears at two-loop order.  For the specific case $r_w = 1.0,\>\>\rho = 1.368$, we find,
  \bea
  Z_{Q\rightarrow Q}^{\overline{\text{MS}},\text{Latt.}}(a \mu, g_0) &=&1 + \frac{g_0^2}{16\pi^2} C_F\left(\frac{8}{3} \log a^2 \mu^2 -1.2795\right)\\
   Z_{G\rightarrow Q}^{\overline{\text{MS}},\text{Latt.}}(a \mu, g_0) &=&- \frac{g_0^2}{16\pi^2} C_F\left(\frac{8}{3} \log a^2 \mu^2 - 2.3760 \right)\\
    Z_{Q\rightarrow G}^{\overline{\text{MS}},\text{Latt.}}(a \mu, g_0) &=&- \frac{g_0^2}{16\pi^2} N_F\left(\frac{2}{3} \log a^2 \mu^2 -0.6522 \right)\\
     Z_{G\rightarrow G}^{\overline{\text{MS}},\text{Latt.}}(a \mu, g_0) &=&1 + \frac{g_0^2}{16\pi^2} \left(\frac{2}{3} N_F \log a^2 \mu^2 - 3.2796 N_c -  1.12484 N_F \right).
  \eea
%
 \section{Summary}\label{conclusions}
In this work we have studied the renormalization and mixing constants for the glue EM tensor operator built from the overlap Dirac derivative for the first time.  These results represent an indispensable piece of a complete calculation of the quark and glue momentum and angular momentum in the nucleon on a quenched 16$^3\>\times\>$24 lattice with three quark masses~\cite{Deka:2013zha}.  There, it was found that reasonable signals were obtained for the glue operator constructed from the overlap Dirac operator.    

The finite contributions to our Z factors reported in the previous section are used to match the lattice results reported in~\cite{Deka:2013zha} to the continuum $\overline{\text{MS}}$ scheme at 2 GeV.  We have commented in previous sections that throughout the course of the calculations we have kept all analytic expressions before a final numerical integration using several python and Mathematica scripts.  Although this allowed us to control all Lorentz and color structures at each stage of the calculation and check explicitly the cancellation of both 1/a and infrared divergences at intermediate stages in the calculation, these benefits came at a cost.  When dealing with the overlap derivative, we found that many intermediate expressions explode in size, requiring intermediate results to be written to disk, slowing down the code substantially.  For future work, we would like to extend our codes to incorporate more complicated lattice actions involving several steps of HYP smearing for the overlap fermion and improved gauge actions.  
\newline
\\
\\
\\
\noindent\textit{Acknowledgments}\newline
\noindent The authors would like to thank G. Von Hippel, S. Capitani, I. Horvath, and X.D. Ji, for their valuable comments and discussions throughout the course of this work.  This research is partially supported by 
 and the U.S. Department of Energy under grant  DE-FG05-84ER40154.  We also thank the Center for Computational Sciences at the University of Kentucky for their financial support of M.G.
   

  \newpage
\appendix
%
%
%
%
\section{Feynman Rules for the Gluon Operator from the Overlap Derivative}\label{appendixA}
In this section we provide details on the derivation of the momentum space Feynman rules for $G_{\mu\nu}$ defined from the Dirac overlap operator.  To start, we define some necessary notation, the form of the Wilson derivative most convenient to these calculations is given by the expressions,
\bea
\hat{X} &=& \sum_\mu \frac{1}{2}\left\{\gamma_\mu (\nabla_\mu^* +\nabla_\mu) - a r_w \nabla_\mu^* \nabla_\mu\right\} -\frac{\rho}{a}\\
\nn\\
\nabla_\mu \psi(x)  &=& \frac{1}{a} \left(U_\mu (x) \psi(x+a\hat{\mu}) - \psi(x)\right)\\
\nabla_\mu^* \psi(x) &=& \frac{1}{a} \left(\psi(x) - U_\mu^\dagger(x-a\hat{\mu}) \psi(x-a\hat{\mu})\right)
\eea
where the gauge-link $U_\mu(x) = \exp(i g_0 a A_\mu(x))$ admits an expansion in the coupling $g_0$.  We denote this order in $g_0$ by giving $\hat{X}$ a subscript, thus $X_0$ corresponds to a zeroth order expansion in $g_0$.  In momentum space, the $X_i$ for ($i=0,1,2$)  are,
\bea
X_0(p) &=& \frac{i}{a}\sum_\mu \gamma_\mu\>\sin ap_\mu + \frac{r_w}{a}\sum_\mu(1-\cos ap_\mu)-\frac{\rho}{a}\\
X_1(p_1,p_2) &=& -g_0\left(i\gamma_\mu \cos\left(\frac{ap_{1}+ap_{2}}{2}\right)_\mu+r_w\sin\left(\frac{ap_1+ap_2}{2}\right)_\mu\right)\\
X_2(p_1,p_2) &=& -\frac{ag_0^2}{2}\left(-i\gamma_\mu \sin\left(\frac{ap_{1}+ap_{2}}{2}\right)_\mu+r_w\cos\left(\frac{ap_1+ap_2}{2}\right)_\mu\right),
\eea
where in these definitions, $p_1$ is the momentum for the incoming fermion, and $p_2$ is the momentum of the outgoing fermion.  We use momentum conservation, $p_2 = p_1 + q$ where $q$ is the momentum of the incoming gluon frequently.  

After following the procedure outlined in section \ref{operators}, we have the following expressions for the first, second and third order Feynman rules for $\text{tr}\>\sigma_{\mu\nu}D_{\text{ov}}(x,x)$
\begin{small}
\bea
\left(\text{tr}\> \sigma_{\mu\nu}D_{\text{ov}}\right)^1& \equiv& G^1_{\mu\nu}(x,x)  = \frac{\rho}{\pi a}\>\text{tr}\>\sigma_{\mu\nu}\int_{-\infty}^\infty d\sigma\sum_y \int_k e^{ikx}\>\hat{\Pi}\left(\sigma^2 X_1 - X_0X_1^\dagger X_0\right)\hat{\Pi}\>e^{-iky}\label{oo1}\\
\nn\\
\nn\\
\left(\text{tr}\> \sigma_{\mu\nu}D_{\text{ov}}\right)^2 &\equiv& G^2_{\mu\nu}(x,x)  = \frac{\rho}{\pi a}\>\text{tr}\>\sigma_{\mu\nu}\int_{-\infty}^\infty d\sigma\sum_y \int_k e^{ikx}\left( \hat{\Pi} \left\{\sigma^2 X_2 - X_0 X_2^\dagger X_0\right\}\hat{\Pi}\right.\nn\\
 &&\left.-\hat{\Pi}\left\{\sigma^2(X_1 X_0^\dagger X_1 + X_1 X_1^\dagger X_0 + X_0 X_1^\dagger X_1)- X_0 X_1^\dagger X_0 X_1^\dagger X_0\right\}\hat{\Pi}^2\right)e^{-iky}\label{oo2}\\
 \nn\\
\left(\text{tr}\> \sigma_{\mu\nu}D_{\text{ov}}\right)^3 &\equiv& G^3_{\mu\nu}(x,x) = \frac{\rho}{\pi a} \>\text{tr}\>\sigma_{\mu\nu}\int_{-\infty}^{\infty}d \sigma\sum_{y}\int_k e^{ikx}\left(\hat{\Pi}\left\{\sigma^2 X_3 -\mathcal{A}\right\}\hat{\Pi} + \hat{\Pi}^2 \left\{\sigma^2 \mathcal{B}+\mathcal{C}\right\}\hat{\Pi}^2\right)e^{-iky}\label{o3}\nn\\,
\eea
\end{small}
where $\hat{\Pi}=\frac{1}{\sigma^2+\Sigma_0}$, $\Sigma_0 = X_0^\dagger X_0$ and $\mathcal{A}, \mathcal{B}$ and $\mathcal{C}$ in $\mathcal{G}_{\mu\nu}^3$ are given by,
\begin{small}
\bea
\mathcal{A} &=& 
X_2 X_0^\dagger X_1 + 
X_2 X_1^\dagger X_0 +
X_1 X_0^\dagger X_2 +
X_1 X_2^\dagger X_0 \nn\\&+&
X_1 X_1^\dagger X_1 +
X_0 X_2^\dagger X_1 +
X_0 X_1^\dagger X_2 +
X_0 X_3^\dagger X_0\\
\nn\\
\mathcal{B} &=& 
X_1 X_0^\dagger X_1 X_0^\dagger X_1 + 
X_1 X_0^\dagger X_1 X_1^\dagger X_0 +
X_1 X_1^\dagger X_0 X_1^\dagger X_0 +
X_0 X_2^\dagger X_0 X_1^\dagger X_0\nn\\ &+&
X_0 X_1^\dagger X_1 X_0^\dagger X_1 +
X_0 X_1^\dagger X_1 X_1^\dagger X_0 +
X_0 X_1^\dagger X_0 X_2^\dagger X_0 +
X_0 X_1^\dagger X_0 X_1^\dagger X_1 \nn\\
&+&\Sigma_0\left(X_2 X_0^\dagger X_1 + 
X_2 X_1^\dagger X_0 +
X_1 X_0^\dagger X_2 +
X_1 X_2^\dagger X_0 + 2X_1 X_1^\dagger X_1+
X_0 X_2^\dagger X_1 +
X_0 X_1^\dagger X_2 \right)\\
\nn\\
\mathcal{C} &=&
\Sigma_0^2\left (X_2 X_0^\dagger X_1 + 
X_2 X_1^\dagger X_0 +
X_1 X_0^\dagger X_2 +
X_1 X_2^\dagger X_0+X_1 X_1^\dagger X_1 +
X_0 X_2^\dagger X_1 +
X_0 X_1^\dagger X_2 
\right) \nn\\
&&+ \Sigma_0\left (X_0 X_2^\dagger X_0 X_1^\dagger X_0 + 
X_0 X_1^\dagger X_0 X_2^\dagger X_0\right )- X_0 X_1^\dagger X_0 X_1^\dagger X_0 X_1^\dagger X_0.
\eea
\end{small}
We shall provide full details for the derivation of the first order result and only sketch the derivation for the second and third orders, since the methods are the same and the intermediate expressions are quite lengthy.  For the third order calculation, we have automated most steps using {\tt FORM}.  

For the first order Feynman rule, we begin by computing the action of $X_0 X_1^\dagger X_0$ on $e^{-i kx} \hat{1}$, and note that the $\sigma^2 X_1$ contribution will not survive the trace as it only contains Lorentz scalar and vector components.  We make use of the general results,
\bea
X e^{-i ky} f(x) &=& e^{-iky} \left\{\sum_\mu \gamma_\mu \left[\tilde{Q}_\mu - \frac{i}{a}s_\mu\right]-r_w\sum_\mu \left[-\frac{1}{a}(1-c_\mu)+\tilde{R}_\mu\right]-\frac{\rho}{a}\right\}f(x)\\
\tilde{Q}_\mu &=& \frac{1}{2}\left(e^{-ik_\mu} \nabla_\mu + e^{ik_\mu}\nabla^*_\mu\right),\>\>\> \tilde{R}_\mu = \frac{1}{2}\left(e^{-ik_\mu}\nabla_\mu - e^{ik_\mu}\nabla^*_\mu\right),\nn\\
s_\mu &=& \sin k_\mu,\>\>\> c_\mu = \cos k_\mu\nn.
\eea
We can now compute the action of various $X_i$ operators on $\hat{1}e^{ikx}$.  For example, for the $\hat{\Pi}$ terms acting on $e^{ikx}\hat{1}$, we can show,
\bea
\hat{\Pi} e^{ikx} &=& \frac{1}{\sigma^2 +X_0^\dagger X_0} e^{ikx} \hat{1} = e^{ikx} \hat{1}\>\frac{1}{\sigma^2 +\omega^2}\nn\\
\nn\\
\omega^2 &=& \frac{1}{a^2}\left(\sum_\mu s_\mu^2 + \sum_\mu\left[r_w(1-c_\mu)-\rho\right]^2 \right),
\eea
where we have used the fact that the derivative terms $\tilde{Q}_\mu$ and $\tilde{R}_\mu$ appearing in $\hat{X}$ acting on $\hat{1}$ vanish, and that one can write the $\hat{\Pi}$ operator as a polynomial in $X^\dagger_0 X_0$.  We are then left with computing the action of $X_0 X^\dagger_1 X_0$ on $\hat{1}e^{ikx}$, calculating each term, 
\bea
X_0 X_1^\dagger X_0 \hat{1} e^{-iky} &=& \hat{1}e^{ipy}\frac{1}{a}\sum_\mu \left(i \gamma_\mu \sin(-k+p)_\mu + r_w(1-\cos(-k+p))_\mu -\frac{\rho}{a}\right)\nn\\
&\times&\>g T^a \left(i \gamma_\rho \cos (-k + p/2)_\rho + r_w \sin (-k + p/2)_\rho\right)\nn\\
&\times&\>\frac{1}{a}\sum_\nu \left(-i \gamma_\nu s_\nu + r_w(1-c_\nu)-\frac{\rho}{a}\right)\\
\nn\\
&\equiv& \hat{1}e^{ipy}\frac{gT^a }{a^2} \sum_{\mu\nu} (\gamma_\mu A^0_\mu + B_\mu^0)\> (\gamma_\rho A^1_\rho + B_\rho^1)\> (\gamma_\nu \bar{A}^0_\nu + \bar{B}_\nu^0).
\eea
In the last line we have used the shorthand,
\bea
A_\mu^0 &=& i \sin(-k+p)_\mu,\>\>\>\>\>\>\> \>\>B_\mu^0 = r_w(1-\cos(-k+p)_\mu)-\rho/a\nn\\
A^1_\rho &=& i\cos(-k+p/2)_\rho,\>\>\>\>\>B_\rho^1 = r_w \sin(-k+p/2)_\rho\nn\\
\bar{A}^0_\nu &=& -i s_\nu,\qquad\qquad\>\>\>\>\>\> \>\>\>\>\bar{B}_\nu^0 = r_w (1-c_\nu) - \rho/a.
\eea
In these expressions, the momentum $k$ is a dummy momentum which is to be integrated and $p$ (with Lorentz index $\rho$, and color index $a$) is the momentum of the incoming gauge field.  At this stage, we compute the trace over Lorentz indices, using the identity,
\bea
\text{tr}\>\sigma_{\mu\nu} \gamma_\alpha \gamma_\beta = 4 i (\delta_{\mu\alpha}\delta_{\nu\beta} - \delta_{\nu\alpha}\delta_{\mu\beta})
\eea
as well as the fact that a trace over an odd number of gamma matrices will vanish.  Performing the trace gives the numerator at first order in the coupling $g$,
\begin{small}
\bea
\mathcal{N}_{\rho;\alpha\beta}&=&\text{tr}\>\sigma_{\alpha\beta}\>\frac{gT^a }{a^2} \sum_{\mu\nu} (\gamma_\mu A^0_\mu + B_\mu^0)\> (\gamma_\rho A^1_\rho + B_\rho^1)\> (\gamma_\nu \bar{A}^0_\nu + \bar{B}_\nu^0)\nn\\
&=&\frac{4 i g T^a}{a^2}\left\{ \sum_\nu A^1_\rho \bar{B}_\nu^0 \left(\delta_{\rho\beta}A_\alpha^0 - \delta_{\rho\alpha}A^0_\beta\right)+B^1_\rho\left(A^0_\alpha \bar{A}_\beta^0 - A^0_\beta \bar{A}^0_\alpha\right)+ \sum_\mu B_\mu^0 A_\rho^1\left(\delta_{\rho\alpha}\bar{A}^0_\beta -\delta_{\rho\beta}\bar{A}_\alpha^0\right)\right\}.\nn\\
\eea
\end{small}
We must still integrate over the $\sigma$-parameter appearing in the various $\hat{\Pi}$ terms of $\mathcal{G}^1_{\alpha\beta}$.  Integrating over $\sigma$ gives,
\bea
\mathcal{G}^{1,a}_{\rho;\alpha\beta}(p)  &=& \frac{\rho}{\pi a}\int_k \>\frac{\mathcal{N}_{\rho;\alpha\beta}(k,p)}{\omega(k)^3}.
\eea
Throughout the course of these calculations, we are not interested in the value of the integral over the dummy momentum $k$, instead we are interested in just the renormalization factor $Z$ which multiplies this operator in momentum space, e.g. we are interested in extracting $Z$ in,
\bea
\int_l\int_k \mathcal{O}(p,k;l) = Z \int_k \mathcal{O}(p,k),
\eea 
where $l$ is the loop momentum of the diagram, and the same integration over dummy $k$ is present on both sides of this equation.  For this reason, we have expanded all $\mathcal{N}(k,p_i)$ numerators and collected all $k$ dependent terms into coefficients multiplying the products of $\sin l$ and $\cos l$.  This is made simpler by the fact that $\omega(k)$ is an even function of $k$ and so all odd functions of $k$ in the numerator can be dropped immediately.  In this way, no integration over the dummy momenta $k$ need be done at any point during the calculations.

The final expression for the zeroth order Feynman rule for the gluon operator is given by the product,
\bea
\mathcal{O}_{\mu\nu}^g(p_1,p_2) &=& \text{tr}_c\>\mathcal{S}\left\{ \mathcal{G}_{\mu\alpha}^1(p_2)\> \mathcal{G}_\alpha^{\nu,1}(p_1)\right\}\nn\\
&=& \text{tr}_c\>\mathcal{S} \left\{\>\frac{\rho^2}{\pi^2 a^2}\int_{k_1,k_2}\frac{\mathcal{N}^a_{\mu_1;\mu\alpha}(k_1,p_1)}{\omega(k_1)^3}\>\frac{\mathcal{N}^b_{\mu_2;\alpha\nu}(k_2,p_2)}{\omega(k_2)^3}\right\},
\eea
where $\mathcal{S}$ reminds us to take the symmetrized and traceless piece of this operator, and tr$_c$ is a trace over the color indices.  Contracting both sides with a light-like vector to project out the symmetrized and traceless piece,
\bea
\Delta\cdot \mathcal{O}^g(p_1,p_2) = \frac{\rho^2\delta^{ab}}{2\pi^2a^2}\>\int_{k_1,k_2}\frac{\Delta\cdot\mathcal{N}_{\mu_1;\alpha}(k_1,p_1)}{\omega(k_1)^3}\>\frac{\Delta\cdot\mathcal{N}_{\mu_2;\alpha}(k_2,p_2)}{\omega(k_2)^3}.
\eea
Here, $p_1$ and $p_2$ are the incoming gauge-field momenta, and it is assumed that all terms odd in $k_1$ and $k_2$ are dropped in the $\mathcal{N}(k_i,p_i)$ numerators.

For the second and third order Feynman rules, all steps of this procedure can be automated.  {\tt FORM} is used to handle all traces and subsequent simplifications.  This is necessary since the third order operator involves many traces over six gamma matrices, for example,
\bea
\text{tr}\>\sigma_{\alpha\beta} \>X_1 X_0^\dagger X_1 X_0^\dagger X_1
\eea
where each $X_i = \gamma_\mu A_\mu^i + B_\mu^i$.  After expanding the traces, and integrating over the $\sigma$ parameter, we construct the full Feynman rule at the desired order from the expansion,
\bea
\mathcal{O}_{\mu\nu}^g = \left(G_{\mu\alpha}^0 + g_0G_{\mu\alpha}^1+g_0^2 G_{\mu\alpha}^2+\ldots\right)\left(G^{0,\alpha}_\nu+g_0G^{1,\alpha}_\nu+g_0^2G^{2,\alpha}_\nu +\ldots\right). 
\eea  
At lowest order, then, we have Feynman rules for two and three external gauge fields respectively, 
\bea
\mathcal{O}^{g,2}_{\mu\nu} &=& g_0^2 G_{\mu\alpha}^1 \> G_{\alpha}^{\nu,1} \\
\nn\\
\mathcal{O}^{g,3}_{\mu\nu} &=& g_0^3 \left(G_{\mu\alpha}^1 \> G_{\alpha}^{\nu,2} +  G_{\mu\alpha}^2 \> G_{\alpha}^{\nu,1}\right)\\
\nn\\
\mathcal{O}^{g,4}_{\mu\nu} &=& g_0^4 \left(G_{\mu\alpha}^1 \> G_{\alpha}^{\nu,3} +  G_{\mu\alpha}^3 \> G_{\alpha}^{\nu,1}+G_{\mu\alpha}^2 \> G_{\alpha}^{\nu,2}\right).
\eea 
We then symmetrize these results over all Lorentz and color indices as well as external momenta.
%
%
%
%
\section{QCD Coupling and Wave Function Renormalization}\label{appendixB}
For completeness, in this section we collect the expressions for $Z_A$ and $Z_\psi$ in the Feynman gauge used to fix the renormalization constants, both are converted under the $\overline{MS}$ scheme. These expressions have been computed elsewhere using the Wilson action, however they serve as an additional check on the accuracy of our codes.  
We write these results in the following form,

\bea
Z_\psi &=& 1-\frac{g_0^2}{16\pi^2}C_F\left(\log a^2 \mu^2 + (F_\psi(r_w)+1)\right)\\
Z_A &=& 1- \frac{g_0^2}{16\pi^2} 
\left( (\frac{5}{3}N_c - \frac{2}{3} N_f)\log a^2 \mu^2 + N_f (F_A(r_w)-\frac{10}{9}) + (B_A(N_c)+\frac{31}{9})\right),
\eea
where 
\bea
B_A(N_c) &=& \frac{2}{9N_c}(7N_c^2 - 12\pi^2)+\frac{2}{9}\pi^2(1+N_c^2)+0.079805N_c 
\eea 
and $F_\psi$, and $F_A$ are evaluated numerically as in Table~[\ref{wfrTable}] for different $r_w$.
\begin{table}[hpb]
\centering
\begin{ruledtabular}
\begin{tabular}{lcccc}
& $r_w$& $F_\psi(r_w)$& $F_A(r_w)$ \\
 \hline\\
  & 0.1 & $5.37037$ &  $1.18502 $ \\
  & 0.2 & $6.13073 $ &  $1.22383  $ \\
 & 0.3 & $7.02470$ &  $1.28534 $ \\ 
 & 0.4 & $7.90649$ &  $1.36776$ \\ 
 & 0.5 & $8.72568$ &  $1.47300$ \\ 
  & 0.6 & $9.47224$ &  $1.60728 $ \\ 
   & 0.7 & $10.1503 $ &  $1.77672$ \\ 
    & 0.8 & $10.7677$ &  $1.97383 $ \\ 
    & 0.9 & $11.3326 $ &  $2.15003 $ \\
    & 1.0 & $11.8524 $ &  $2.16850 $ 
\end{tabular}
\end{ruledtabular}
\caption{Table for the finite contributions to the wave function renormalization constants $Z_\psi$, $Z_A$, and $Z_g$ used to renormalize the quark and gluon angular momentum operators.}
\label{wfrTable}
\end{table}
\begin{figure}
\begin{center}
\includegraphics[scale=0.47]{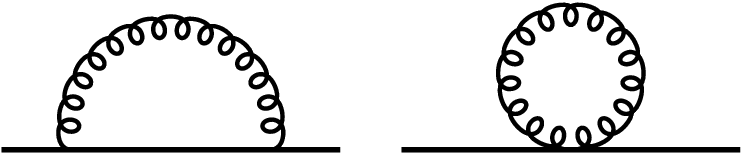}
\caption{Feynman diagrams for the calculation of $Z_{\psi}$.}
\label{quark_wfr}
\end{center}
\end{figure}
\begin{figure}
\begin{center}
\includegraphics[scale=0.60]{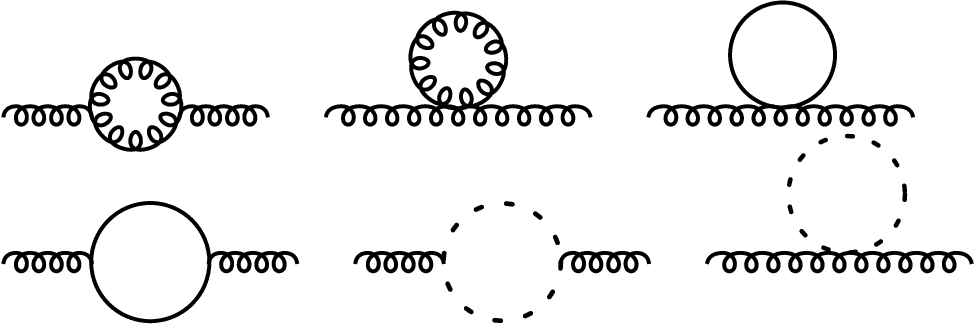}
\caption{Feynman diagrams for the calculation of $Z_{A}$. The straight, dashed and curly lines are the quark, ghost and gluon lines respectively.}
\label{gluon_wfr}
\end{center}
\end{figure}
%
%
%
%

\section{QCD Vertices and Operator Feynman Rules}\label{appendixD}
In this section we collect the various Feynman rules used during the course of the calculations.  For the QCD action, the fermion and gluon propagators take the form,  
\begin{small}
\begin{eqnarray}
  \label{eq:QCD3}
  \parbox{40mm}{
    \begin{center}
      \includegraphics[scale=0.5]{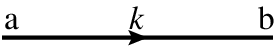}
    \end{center}
  }
&&a \delta_{ab}a\frac{-i\sum_\mu \gamma_\mu \sin ak_\mu  + 2r_w\sum_\mu \sin^2 ak_\mu/2+am_q}{\sum_\mu \sin^2 ak_\mu +\left(2r_w\sum_\mu \sin^2 ak_\mu/2+m_q\right)^2}\\
  \label{eq:QCD3}
  \parbox{40mm}{
    \begin{center}
      \includegraphics[scale=0.5]{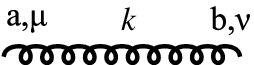}
    \end{center}
  }
&& \frac{\delta_{ab}}{4/a^2\sum_\rho \sin^2 k_\rho/2}\left\{\delta_{\mu\nu}-(1-\alpha)\frac{\sin \frac{a k_\mu}{2}\>\sin \frac{a k_\nu}{2}}{\sum_\rho\sin^2\frac{a k_\rho}{2}}\right\}\\
%
%
  \label{eq:QCD4}
  \parbox{40mm}{
    \begin{center}
      \includegraphics[scale=0.5]{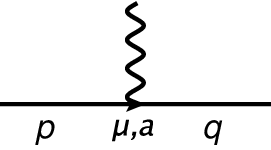}
    \end{center}
  }
&& -g_0T^a\left(i \gamma_\mu\cos \frac{ap_\mu+aq_\mu}{2}+r_w\sin\frac{ap_\mu+aq_\mu}{2}\right)\\
%
%
  \label{eq:QCD5}
  \parbox{40mm}{
    \begin{center}
      \includegraphics[scale=0.5]{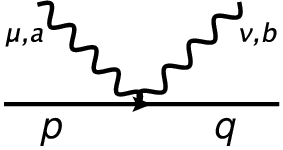}
    \end{center}
  }
&& -\frac{ag_0^2\left\{T^a,T^b\right\}}{2}\delta_{\mu\nu}\left(-i\gamma_\mu\sin \frac{ap_\mu+aq_\mu}{2}+r_w\cos\frac{ap_\mu+aq_\mu}{2}\right)
\end{eqnarray}
\end{small}
%
%
\begin{small}
\begin{eqnarray}
  \label{eq:QOP1}
  \parbox{40mm}{
    \begin{center}
      \includegraphics[scale=0.5]{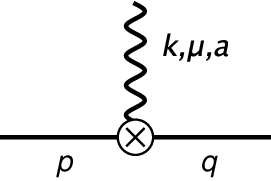}
    \end{center}
  }
&& \frac{ig_0}{2}\sum_{\mu,\nu}T^a \left( \gamma_\nu\cos\frac{ap_\mu+aq_\mu}{2} +  \gamma_\mu \cos \frac{ap_\nu+aq_\nu}{2}\right)
\end{eqnarray}
\end{small}

\bibliographystyle{h-physrev3.bst}
\bibliography{reference.bib}
\end{document}